\documentclass[conference]{IEEEtran}
\IEEEoverridecommandlockouts

\usepackage{cite}
\usepackage{amsmath,amssymb,amsfonts}
\usepackage{algorithmic}
\usepackage{graphicx}
\usepackage{textcomp}
\usepackage{xcolor}
\usepackage{hyperref}
\usepackage{listings}
\usepackage{pifont}
\usepackage{xcolor}
\usepackage[font=normalsize]{caption}

\newcommand{\cmark}{\textcolor{green!60!black}{\ding{51}}} 
\newcommand{\xmark}{\textcolor{red!70!black}{\ding{55}}}   
\newcommand{\pmark}{$\triangle$}
\def\BibTeX{{\rm B\kern-.05em{\sc i\kern-.025em b}\kern-.08em
    T\kern-.1667em\lower.7ex\hbox{E}\kern-.125emX}}

\definecolor{backgroundColour}{HTML}{F8F8F8}
\definecolor{keywordclr}{HTML}{AA00FF}
\definecolor{commentclr}{HTML}{649696}
\definecolor{stringsclr}{HTML}{11A579}
\definecolor{fnctionclr}{rgb}{0.467, 0, 0.533}
\definecolor{builtinclr}{rgb}{0.35, 0, 0.533}
\definecolor{symbolsclr}{rgb}{0.5, 0.25, 0.25}   
\definecolor{numbersclr}{rgb}{0.8, 0.2, 0}
\definecolor{bckgrndclr}{rgb}{0.91, 0.95, 0.95}
\lstdefinestyle{PythonStyle}{
    language=Python,
    backgroundcolor=\color{backgroundColour},
    keywordstyle=\color{keywordclr},
    stringstyle=\color{stringsclr},
    commentstyle=\color{commentclr},
    upquote=true,
    basicstyle=\ttfamily\linespread{0.9}\footnotesize,
    breakatwhitespace=false,
    breaklines=true,
    captionpos=b,
    keepspaces=true,
    numbers=left,
    numbersep=5pt,
    numberstyle=\color{commentclr}\ttfamily\tiny,
    showspaces=false,
    showstringspaces=false,
    showtabs=false,
    tabsize=2,
    xleftmargin=2.25em,
    frame=single,
    framexleftmargin=1.25em,
    morekeywords={assert,with,as,None}
}
\begin{document}

\title{Experiences Building Enterprise-Level Privacy-Preserving Federated Learning to Power AI for Science}

\author{\IEEEauthorblockN{
Zilinghan Li\IEEEauthorrefmark{1}\IEEEauthorrefmark{2},
Aditya Sinha\IEEEauthorrefmark{1}\IEEEauthorrefmark{2}\IEEEauthorrefmark{3},
Yijiang Li\IEEEauthorrefmark{4},
Kyle Chard\IEEEauthorrefmark{1}\IEEEauthorrefmark{5},
Kibaek Kim\IEEEauthorrefmark{4}\IEEEauthorrefmark{6},
Ravi Madduri\IEEEauthorrefmark{1}\IEEEauthorrefmark{6}
}
\IEEEauthorblockA{
\IEEEauthorrefmark{1}Data Science and Learning Division, Argonne National Laboratory \\
\IEEEauthorrefmark{2}Center for AI Innovation, National Center for Supercomputing Applications \\
\IEEEauthorrefmark{3}Siebel School of Computing and Data Science, University of Illinois at Urbana-Champaign \\
\IEEEauthorrefmark{4}Mathematics and Computer Science Division, Argonne National Laboratory \\
\IEEEauthorrefmark{5}Department of Computer Science, The University of Chicago \\
\IEEEauthorrefmark{6}Consortium for Advanced Science and Engineering, The University of Chicago \\
}
\IEEEauthorblockA{
\{zilinghan.li, yijiang.li, kimk, madduri\}@anl.gov, aditya47@illinois.edu, chard@uchicago.edu
}
}

\maketitle

\begin{abstract}
Federated learning (FL) is a promising approach to enabling collaborative model training without centralized data sharing, a crucial requirement in scientific domains where data privacy, ownership, and compliance constraints are critical. However, building user-friendly enterprise-level FL frameworks that are both scalable and privacy-preserving remains challenging, especially when bridging the gap between local prototyping and distributed deployment across heterogeneous client computing infrastructures. In this paper, based on our experiences building the Advanced Privacy-Preserving Federated Learning (APPFL) framework, we present our vision for an enterprise-grade, privacy-preserving FL framework designed to scale seamlessly across computing environments. We identify several key capabilities that such a framework must provide: (1) Scalable local simulation and prototyping to accelerate experimentation and algorithm design; (2) seamless transition from simulation to deployment; (3) distributed deployment across diverse, real-world infrastructures, from personal devices to cloud clusters and HPC systems;  (4) multi-level abstractions that balance ease of use and research flexibility; and (5) comprehensive privacy and security through techniques such as differential privacy, secure aggregation, robust authentication, and confidential computing. We further discuss architectural designs to realize these goals. This framework aims to bridge the gap between research prototypes and enterprise-scale deployment, enabling scalable, reliable, and privacy-preserving AI for science.
\end{abstract}


\section{Introduction}

\begin{figure*}[h]
\centerline{\includegraphics[width=\linewidth]{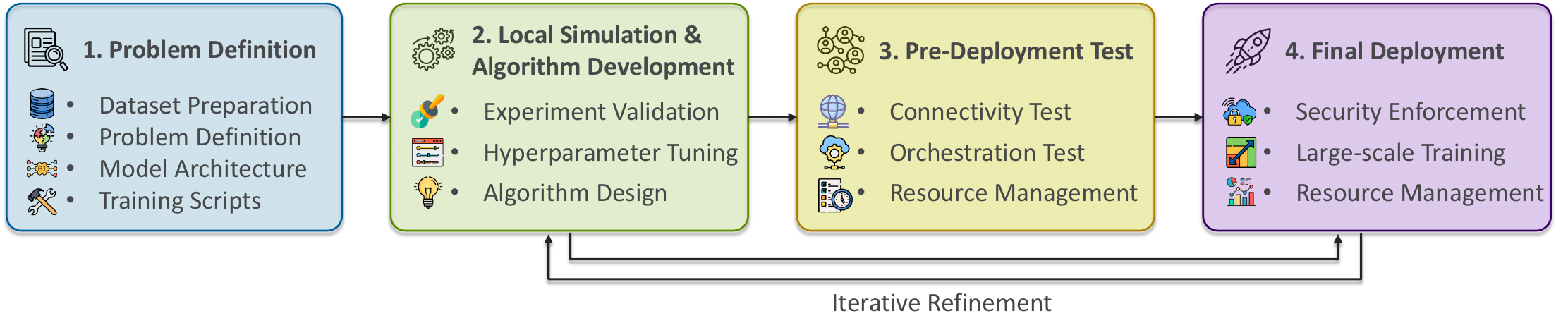}}
\caption{\textbf{A typical end-to-end workflow for leveraging federated learning to train AI models.} The workflow begins with the problem definition phase, where practitioners formulate the learning objective and design the model based on the available data, similar to conventional centralized training. This is followed by local simulation and algorithm development, in which practitioners verify training logic, tune hyperparameters, and prototype aggregation strategies in a simulated federated environment. Next, during the pre-deployment testing stage, the workflow is validated across distributed computing resources to ensure connectivity and configuration consistency. Finally, in the full deployment and training phase, large-scale federated training is executed across heterogeneous clients. Practitioners may iteratively revisit the simulation phase to refine configurations and improve model performance.}
\label{fig:workflow}
\end{figure*}

Artificial intelligence (AI) is rapidly transforming scientific discovery by enabling data-driven modeling and prediction at unprecedented scales. AI has already had transformative impacts in domains from drug discovery \cite{blanco2023role} and genomics \cite{theodoris2023transfer,avsec2025alphagenome} to climate modeling \cite{nguyen2024scaling} and materials science \cite{jablonka202314}. Scientific workflows increasingly  depend on AI; however, due to privacy concerns, intellectual property restrictions, and regulatory compliance requirements, scientific data are often considered sensitive and thus cannot be easily shared. 
These factors make centralized model training difficult, or even infeasible, in many real-world scientific applications. 

Federated Learning (FL), a decentralized machine learning paradigm, addresses these challenges by enabling collaborative model training across distributed data sources.
\cite{konevcny2016federated,mcmahan2017communication}. In FL, models are trained locally where the data reside, and only the model updates are shared among different data holders for model weight aggregation, thus enabling collaborative learning without centralized data sharing. 
The decentralized data paradigm of FL makes it particularly appealing for building AI models for science, where scientific datasets are often geographically distributed across laboratories, research facilities, and computing infrastructures. However, requirements for distributed execution of training and aggregation tasks across heterogeneous computing infrastructure while meeting stringent data security requirements pose significant challenges to deploying FL workflows in practice.
Therefore, there is a strong need for an FL framework that addresses these challenges and simplifies deployment for both researchers and practitioners. 

We propose a vision for an enterprise-level, privacy-preserving FL framework that scales across heterogeneous computing environments. The vision is based on our years of experiences building the Advanced Privacy-Preserving Federated Learning (APPFL) framework \cite{ryu2022appfl,li2025advances}, which has already been adopted to develop federated solutions for various scientific domains, such as biomedicine \cite{hoang2025enabling}, smart grid \cite{bose2024privacy}, and astrophysics \cite{patel2025radio}. The envisioned framework aims to provide a unified and secure infrastructure that seamlessly supports the entire FL lifecycle, from experiment simulation and algorithm development to production-scale deployment, while ensuring robust privacy and security guarantees. Specifically, we identify several key capabilities that such a framework must provide:

\begin{itemize}
    \item Scalable local simulation and prototyping, to accelerate experimentation and algorithm design through efficient and convenient large-scale simulation of federated experiment settings.
    \item Seamless transition from simulation to deployment, minimizing code and configuration changes between local simulation and distributed deployment modes.
    \item Real-world distributed and secure deployment support across heterogeneous  computing infrastructures, including cloud clusters, HPC systems, and personal devices.
    \item Multi-level abstractions that support both applied-users seeking ease of use and researchers requiring algorithmic flexibility.
    \item Comprehensive security and support for privacy-preservation, integrating differential privacy, secure aggregation, homomorphic encryption, secure multi-party computation, robust authentication, and confidential computing.
    \item Ability to easily create secure federations across organizations while evaluating dataset distributions and configuration of different hyper-parameters thus lowering the overall barrier to entry to adopt FL. 
\end{itemize}

Building such a framework requires co-designing FL systems with modern distributed computing technologies. In particular, we envision an architectural design that decouples the core FL logic from complex communication and resource orchestration layers, leveraging tools such as Kubernetes or Ray \cite{moritz2018ray} for cloud orchestration, and Parsl \cite{babuji2019parsl} or Globus Compute \cite{ananthakrishnan2024establishing} for managing workloads on personal devices and HPC systems. The design also facilitates seamless user customization for algorithm design and can be extended to a hosted service model for further lowering the barrier to entry for FL adoption. Ultimately, the proposed framework vision aims to bridge the gap between research prototypes and enterprise-scale deployments, enabling accessible, scalable, reliable, and privacy-preserving AI for science.

\section{Federated Experiment Workflow}
To better identify essential capabilities of a comprehensive enterprise-level FL framework, it is necessary to first understand the typical end-to-end workflow that FL practitioners follow when training AI models. As illustrated in Figure~\ref{fig:workflow}, FL practitioners follow a multi-stage process that bridges problem formulation, experimental simulation, and real-world deployment \cite{kairouz2021advances}. While the specific steps may vary depending on practitioners and use cases, a comprehensive FL workflow generally follows the stages outlined in the subsections below. For example, applied FL users may traverse the entire workflow to train models collaboratively, whereas FL researchers may focus primarily on the simulation and algorithm development stages. Collectively, these stages capture the full lifecycle of an FL workflow and help us highlight the practical challenges that motivate the development of an enterprise-level FL framework.

\subsection{Problem Definition} 
The first step in an FL workflow is to clearly define the problem to be solved and its scientific or application context. Given a set of distributed data sources, practitioners must determine a meaningful learning objective, identify an appropriate AI model architecture, understand data characteristics (AI-readiness) \cite{hiniduma2025cadre},  and prepare the data pre-processing pipeline and training scripts. Establishing a consistent data schema and model interface is crucial to ensuring interoperability across distributed and isolated client environments. At this stage, practitioners may also prepare or collect openly available, non-sensitive data for the defined problem for model building, validation and testing purposes. This stage is similar to developing an AI model using a centralized dataset, serving as the conceptual foundation for subsequent FL stages.

\subsection{Local Simulation and Algorithm Development} Before launching a real-world distributed deployment, practitioners typically conduct local simulation and prototyping to verify that the problem can be effectively trained in a federated setting. This phase often involves running simulated FL experiments on a single machine or a local cluster, using either serial or parallel simulation of multiple clients, depending on the amount of available resources. This simulation phase is essential for algorithmic development, debugging, and reproducibility, providing a controlled environment before moving to distributed deployment. Through this process, practitioners might:

\begin{itemize}
    \item Validate that their client training and server aggregation logic can operate correctly in a federated setting.
    \item Explore and tune hyper-parameters such as learning rates, local epochs, or aggregation frequencies on open federated datasets or heterogeneously partitioned local datasets.
    \item Design and test new algorithms, such as aggregation strategies, compression techniques, or privacy mechanisms for certain application use cases.
\end{itemize}

\subsection{Pre-deployment Testing} Once the algorithm and experiment settings have been validated through local simulation, the next stage involves preparing for distributed deployment, which serves as the bridge between local simulation and full-scale execution. Here, a group of real clients comes together to verify system connectivity, configuration consistency, workflow orchestration, and resource management correctness. By validating these practical aspects early, practitioners can detect and resolve potential issues in distributed execution before initiating large-scale experiments. This initial test stage typically uses small datasets (often synthetic or publicly available) and minimal computing resources to ensure that:

\begin{itemize}
    \item Secure connections between the server and clients are properly established.
    \item Data loading, local training, and update aggregation function as expected.
    \item Resource schedulers (e.g., Kubernetes or HPC workflow management tools) correctly handle task execution across various client computing resources.
\end{itemize}

\subsection{Final Deployment} In the final stage, the FL workflow is deployed in a distributed environment with the actual participating clients, private datasets, and heterogeneous computing resources. This is where large-scale training, evaluation, and model refinement take place. At this point, the framework must coordinate clients across diverse infrastructures, ranging from cloud clusters and HPC systems to on-premise servers and personal devices, while maintaining strict privacy and security guarantees. In this stage, the FL system may need to manage challenges such as asynchronous client participation and failure recovery, heterogeneity in computational power and network bandwidth, and secure aggregation and privacy enforcement at scale. If the resulting global model does not meet the performance expectation, users may iteratively revisit earlier stages, most often the local simulation for algorithm tuning or aggregation strategy refinement, to optimize the training outcome. This iterative feedback loop between prototyping and deployment ensures continuous improvement of large-scale collaborative training.

\section{Framework Capabilities} 
Building on the FL workflow described in the previous section, we now identify and elaborate on the core capabilities that an enterprise-level FL framework should provide to support each stage of the workflow. These capabilities are designed to address the practical and technical challenges encountered throughout the entire FL lifecycle. By aligning framework design with the natural progression of an FL experiment, we aim to enable a unified system that is scalable, flexible, and privacy-preserving, while remaining accessible and user-friendly to both advance users and algorithm developers.

\subsection{Scalable Local Simulation and Prototyping}
In the local simulation and algorithm development stage, practitioners start to launch the model training in the federated setting with multiple clients using one or multiple processes on a single machine or cluster. It is one of the most critical phases in the FL workflow, as it is where most of the feasibility verification, experimental configuration, and algorithm design happen. Local simulation shortens the development cycle and enables faster, more flexible experimentation for distributed learning paradigms. This phase is essential for rapid iteration, debugging, and algorithmic innovation, allowing practitioners to evaluate FL algorithm behaviors without deploying complex multi-site infrastructures. Therefore, an enterprise-level FL framework must provide robust built-in support for scalable local simulation to accelerate development. However, supporting large-scale local simulation presents several system-level challenges. Simulating multiple clients on a single machine or local cluster can be resource-intensive, especially when working with large models and datasets. Efficiently managing concurrency, GPU sharing, and communication emulation is nontrivial. Furthermore, existing FL frameworks often require practitioners to manually configure client simulations, making it difficult to scale beyond a small number of clients or to reproduce results consistently across environments. To address these challenges, a well-designed FL framework should include the following key features:

\begin{itemize}
    \item Automated orchestration of large-scale simulated clients: Enable practitioners to seamlessly scale from a few to hundreds or even thousands of clients, potentially by implementing virtual clients, to test federated training of various scales.
    \item Parallel execution and resource management: Efficiently utilize multi-core CPUs and multi-GPU systems to support both serial and parallel local simulation, ensuring optimal resource utilization and improving simulation efficiency.
    \item Configurable data partitioning utilities: Provide flexible tools to emulate diverse, non-IID data distributions that reflect real-world FL conditions, enabling convenient and realistic local experimentation.
\end{itemize}

\subsection{Seamless Transition from Simulation to Deployment}
After the local simulation and algorithm development phase, practitioners typically need to migrate their experiments to a distributed environment for initial testing across real computing resources. This transition phase, which bridges local prototyping and large-scale deployment, is critical for verifying connectivity, compatibility, and stability across diverse infrastructure such as cloud clusters, HPC systems, and institutional servers. At this stage, practitioners may also determine the appropriate workload allocation for each client and estimate the required computational resources for full-scale execution. A well-designed FL framework should therefore enable practitioners to seamlessly transition from local simulation to distributed pre-deployment testing with minimal engineering overhead. However, in many existing FL frameworks, the simulation and deployment environments are largely decoupled. Practitioners are often required to make extensive and non-trivial modifications to experiment scripts, reconfigure client execution logic, or manually set up new communication protocols. Such discrepancies might significantly increase deployment complexity, slow down overall experiment progress, and may even cause inconsistencies between simulation results and real-world performance. These challenges are particularly pronounced in scientific domains, where domain experts, often with limited systems or infrastructure expertise, seek to leverage FL for building AI for Science applications.

To address this challenges, an enterprise-level FL framework should provide a unified development and deployment interface that makes sure that the same experiment design and definition should be executable both locally and in distributed settings with only minimal adjustments. This design philosophy not only accelerate the transition from simulation to deployment but also reduces human error during setup and promotes reproducibility. More specifically, the framework should provide the following features:

\begin{itemize}
\item Consistent APIs and configuration schemas: Maintain identical experiment definitions and interfaces for local and distributed runs to avoid major code refactoring during migration.
\item Configurable runtime backends: Allow practitioners to switch between local simulation, cloud computing orchestration, and HPC cluster schedulers through standardized configuration files or command line arguments.
\item Automated environment packaging and deployment: Provide tools for containerization, dependency tracking, and deployment scripting to simplify launching pre-deployment tests.
\item Unified monitoring and experiment tracking: Ensure that metrics, logs, and intermediate results collected during simulation are seamlessly integrated with those from pre-deployment runs for easy comparison and debugging in distributed settings.
\end{itemize}

\subsection{Distributed Deployment Across Heterogeneous Infrastructures}
After pre-deployment testing verifies the basic connectivity and workflow integrity, the next step involves large-scale distributed deployment across clients operating on heterogeneous computing infrastructures. While the FL research community has made substantial progress in algorithmic design to address the challenges such as convergence, efficiency, and communication cost, practical deployment across heterogeneous environments remains a major barrier to scaling FL for many scientific applications. 

Heterogeneity in participating clients spans many dimensions, such as computational capabilities, resource scheduling policies, software maturity, and security constraints.  For example, among common hardware accelerators such as NVIDIA, AMD, and Intel systems, the collective communication libraries that serve as the backbone for both intra- and inter-node communications are different: NVIDIA systems rely on NCCL for optimized GPU-to-GPU communication, AMD systems use RCCL, and Intel systems employ oneCCL. These libraries have different version requirements, API nuances, and performance characteristics. Fundamental operations like broadcasting data from a root rank to all workers can exhibit performance disparities spanning multiple orders of magnitude across different hardware platforms. Such performance differences necessitate platform-specific communication approaches that deviate from one-size-fits-all implementations, where operations that perform adequately on some hardware become severe bottlenecks on other systems. 

Another example of heterogeneity is the diverse computing resource providers for modern scientific computing. These include common cloud platforms such as Amazon Web Services, Google Cloud Platform, and Microsoft Azure, as well as shared national research infrastructures for supporting scientific AI applications. For example, the National Artificial Intelligence Research Resource (NAIRR) Pilot \cite{NSF_NAIRR_Pilot_2024} has already facilitated many AI for Science projects by providing access to various computing resources. Similarly, the forthcoming American Science Cloud (AmSC) initiative \cite{DOE_AmSC_2025} aims to interconnect U.S. Department of Energy laboratories, national computing facilities, and academic institutions through a unified cloud-based research ecosystem to enable scalable data-intensive research. To effectively operate across such diverse and federated computing environments, the FL framework should be designed with a modular approach with well-defined interfaces and separation of concerns that enable seamless integration with different computational, identity and access management, co-scheduling interfaces that allow for large-scale, secure federated learning workloads across multiple infrastructures. 




A well-designed FL framework should shield practitioners from platform-specific complexity, allowing domain scientists and researchers to focus on their models, data, and scientific questions rather than becoming experts in hardware-specific tuning, compatibility matrices, or system configuration optimization. Specifically, to support distributed deployment across heterogeneous infrastructures, an effective FL framework should provide the following capabilities:

\begin{itemize}
\item \text{Multi-environment orchestration:} Integrate cloud-native systems with HPC schedulers through pluggable backends and unified job submission interfaces.
\item \text{Cross-provider interoperability:} Enable consistent execution across various administrative domains, from commercial clouds and HPC centers to personal devices, while ensuring compliance with regulatory security policies.
\item \text{Automatic environment detection and adaptation:} Detect the underlying hardware and software environment, query available accelerators and library versions, and simplify the selection of optimized communication backends, parallelism strategies, and configuration parameters.
\item \text{Resource-aware scheduling:} Dynamically allocate workloads based on client capabilities and data sizes to maximize overall training efficiency.
\item \text{Fault tolerance and elasticity:} Support asynchronous participation and recovery mechanisms to handle intermittent connectivity and variable availability across clients.
\item \text{Secure and reliable communication:} Provide standardized and encrypted channels (e.g., gRPC over TLS, or MPI-based collectives for HPC) to ensure data integrity and privacy across distributed networks.
\end{itemize}

\subsection{Hierarchical Abstractions for Usability and Extensibility}

In practice, practitioners of an FL framework might have diverse objectives, levels of expertise, and expectations. While many domain users may rely on existing popular FL strategies provided by the framework for their applications, other FL researchers may need to design customized training or aggregation algorithms tailored to their data characteristics, research goals, or performance requirements, particularly when the trained models from the final deployment stage fall short of desired accuracy or generalization using existing algorithms. In addition, algorithmic developers and systems researchers often focus on advancing FL methodology itself. Therefore, a robust enterprise-level FL framework must support both domain users seeking usability and developers pursuing extensibility.

To achieve this, the framework should adopt a hierarchical abstraction design that provides multiple levels of control and interfaces. At the high level, domain users should be able to define FL experiments through simple configuration files or declarative APIs. Users can specify model architectures, datasets, and basic parameters without requiring knowledge of underlying framework implementation details. This enables quick setup and execution of FL experiments, reducing the engineering barrier for practitioners with limited distributed systems experience. At the low level, the framework should expose well-defined and modular interfaces for core components such as client training logic, aggregation algorithms, communication backends, and privacy mechanisms. This modularity allows advanced researchers to easily extend or replace specific modules to explore new research ideas while maintaining overall compatibility and stability. By supporting both levels of abstraction, the framework can simultaneously foster accessibility for domain users and flexibility for researchers, enabling innovation and usability within the same unified system.

\subsection{Comprehensive Privacy and Security Guarantees}

One of the key motivations for leveraging FL is to preserve the privacy of sensitive training data. Although raw data are never directly shared among participating clients, privacy leakage still remains possible. For example, gradient inversion can potentially reconstruct sensitive training data samples from shared updates, thus undermining data confidentiality \cite{zhao2020idlg,geiping2020inverting,yin2021see,hatamizadeh2023gradient}. Additionally, due to limited control and visibility in the clients participating in FL, the experiment could be vulnerable to Byzantine clients who may upload corrupted models or use poisoned data for training, significantly degrading the trained model performance \cite{shejwalkar2021manipulating,bagdasaryan2020backdoor,tolpegin2020data,fang2020local}. Therefore, privacy and security are among the most fundamental design requirements of an FL framework, which must incorporate a comprehensive suite of mechanisms operating on multiple layers of the system stack.

The most common approaches to preserving privacy in FL include Differential Privacy (DP), which perturbs model updates with carefully calibrated noise to provide quantifiable privacy guarantees \cite{abadi2016deep,bonawitz2017practical}, and Secure Aggregation (SecAgg), which enables the server to aggregate encrypted client updates without accessing individual contributions \cite{fereidooni2021safelearn}. These methods effectively protect against reconstruction attacks while maintaining the utility of global model. Beyond these techniques, Homomorphic Encryption (HE) \cite{aono2017privacy,hardy2017private} and Secure Multi-Party Computation (SMC) \cite{mohassel2017secureml} can be employed for stronger cryptographic protection, though often with higher computational overhead. 

In addition to algorithmic defenses, system-level protections are also essential. Confidential computing techniques, such as Intel SGX and AWS Nitro Enclaves, enables secure code execution within Trusted Execution Environments (TEEs) that isolate sensitive computing from untrusted system components \cite{mo2021ppfl}. Complementing this, secure container technologies such as NVIDIA confidential containers provide an additional layer of software-based isolation. Secure containers can sandbox FL workloads within lightweight virtualized environments, ensuring that even if the host OS or other workloads are compromised, access to model parameters or intermediate data remains restricted. Secure containers are particularly valuable in hybrid cloud–HPC federations, where full TEE deployment may not be feasible on all nodes. Finally, the framework should support end-to-end encryption for all communication channels and enforce robust authentication mechanisms to verify the identity and integrity of participating clients, ensuring that only authorized clients and servers can join the federation \cite{li2023appflx}. 


\section{Architectural Framework Design}
\begin{figure}[htbp]
    \centering
    \includegraphics[width=\linewidth]{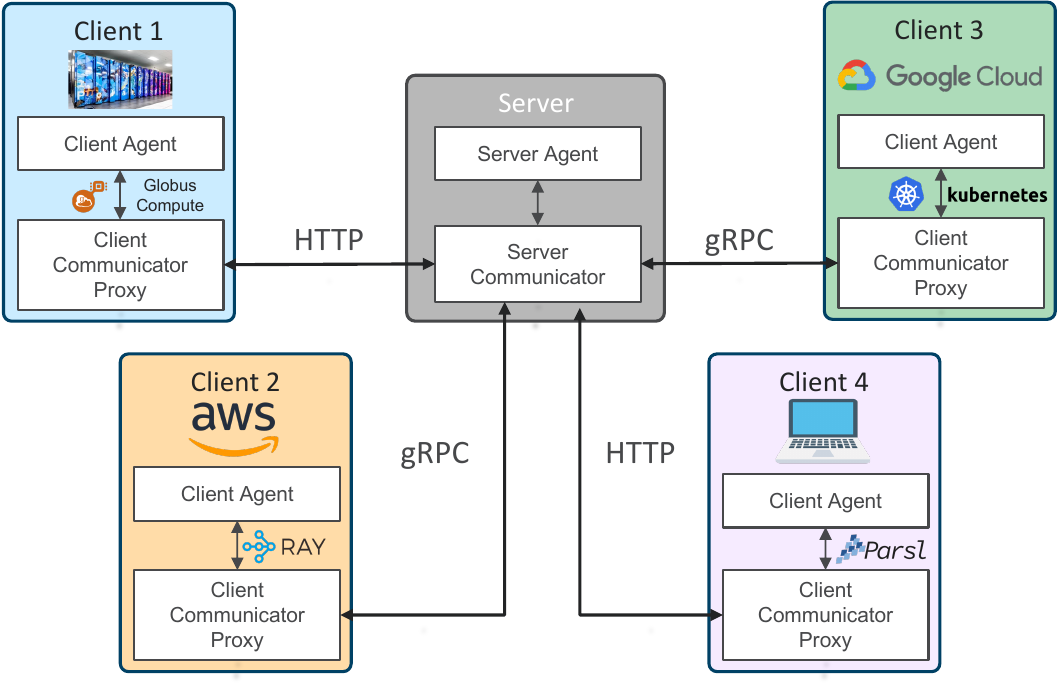}
    \caption{\textbf{Architectural design of the proposed FL framework.} The central server coordinates multiple heterogeneous clients deployed across diverse computing environments - including HPC clusters, cloud computing (e.g, AWS, Google Cloud), and personal devices. Each client agent performs local model training with its platform-specific resource management ecosystem (e.g., Ray and Kubernetes for cloud computing, Globus Compute and Parsl for HPC or personal devices), while the server agent manages global aggregation and orchestration. Communication between clients and the server occurs over HTTP or gRPC, enabling efficient and platform-agnostic interaction.}
    \label{fig:architecture_diagram}
\end{figure}
To realize the vision outlined in this paper, we propose a modular, extensible, and scalable architectural design shown in Figure~\ref{fig:architecture_diagram}. The core philosophy of this design is the separation of concerns, decoupling the fundamental FL logic from the complexities of network communication and infrastructure management. This separation is key to achieving seamless transition from local simulation to distributed deployment and providing the hierarchical abstractions necessary for both ease of use and research flexibility. Our proposed architecture is composed of four primary components, organized into a server-side and client-side stack, communicating via protocols such as gRPC or HTTP/HTTPS.

\subsection{Core Components}
\textit{Server Agent}: This component is the central orchestrator of the FL process. Its primary responsibilities include: (1) maintaining the state of the global model, (2) executing a synchronous or asynchronous model aggregation strategy after each training round (e.g., FedAvg \cite{mcmahan2017communication}, FedProx \cite{li2020federated}, and FedCompass \cite{li2023fedcompass}), (3) managing the entire FL lifecycle, 
(4) coordinating the selection of clients for each round, and (5) running server-side privacy-preserving algorithms. This component encapsulates the core server-side algorithmic logic of the federation.
    
\textit{Server Communicator}: This is a lightweight component that acts as the dedicated network interface for the server agent. It abstracts the complexities of the underlying communication protocol. Its sole responsibility is to handle all network I/O: listening for incoming client connections, distributing the global model to participating clients, receiving model updates from clients, and passing these updates securely to the server agent for aggregation. Different implementations may be developed depending on the execution mode. For example, different communicators might be needed for parallel local simulations and distributed deployments, while for single-processor simulations, no communicator is needed. 
By decoupling communication from the core orchestration and aggregation logic, the server agent can operate independently of the running mode and network topology, enhancing modularity and seamless transition between simulation and deployment. 

\textit{Client Agent}: This component is the training worker deployed on each client machine. It is responsible for executing the local training task. Upon receiving the global model from the server, the client agent conducts the local training loop on the client's private data for a specified number of steps/epochs. It encapsulates all client-side logic, such as data loading, model training, and evaluation. This is also the component where client-side privacy mechanisms, such as applying differential privacy noise to gradients, would be implemented.
    
\textit{Client Communication Proxy}: Analogous to its server-side counterpart, the client communication proxy manages all network interactions on behalf of the client agent. It is responsible for initiating contact with the server, requesting the initial global model, transmitting the locally trained model updates back to the server, and receiving the updated global model for subsequent rounds. In addition to that, in a real-world deployment, this component also needs to handle the instantiation and lifecycle management of the client agent across heterogeneous resources by using various workflow management tools, such as Kubernetes or Ray \cite{moritz2018ray} for cloud-based clusters, and Globus Compute \cite{chard2020funcx} or Parsl \cite{babuji2019parsl} for coordinating training workloads on HPC systems or personal devices. The proxy decouples client training logic with complex communication and resource coordination, making it simple for simulation and deployment transitions. 

\subsection{Extensibility}
A critical requirement for a FL framework that is both user-oriented and researcher-oriented is extensibility. To empower researchers to innovate without understanding all the framework implementation details and modifying its core codebase, we propose a possible implementation solution with a hook-based event-driven architecture. This approach exposes well-defined lifecycle events at critical stages of the FL process on both the server and client sides.

Practitioners can register custom Python functions (callbacks) to these hooks, which are executed when the corresponding event is triggered. To enable stateful and intelligent customizations, each callback function is passed a \texttt{context} object containing the current state of the system (e.g., current round number, global model, client-side metrics, etc.).

We envision a series of hooks such as:
\begin{itemize}
    \item \text{Server-side hooks:}
    \begin{itemize}
        \item \texttt{on\_server\_start} \item \texttt{before\_client\_selection} \item \texttt{before\_aggregation} \item \texttt{after\_aggregation} \item \texttt{on\_experiment\_end}
    \end{itemize}
    \item \text{Client-side hooks:}
    \begin{itemize}
        \item \texttt{on\_client\_start} \item \texttt{before\_local\_train} \item \texttt{after\_local\_train} \item \texttt{before\_model\_upload}
    \end{itemize}
\end{itemize}

For example, a researcher who wants to evaluate the performance of the local model on each client's private test dataset can do so without altering the core training loop as shown in Listing~\ref{lst:example1}. They would define a function decorated with \texttt{@on\_event(`after\_local\_train')} that takes \texttt{server\_context} and \texttt{client\_context} as arguments. Within this function, the local model and test data can be accessed from \texttt{client\_context} (e.g., \texttt{client\_context.model}, \texttt{client\_context.data.test\_loader}). The researcher can then evaluate the model, compute metrics such as loss and accuracy, and record them in the \texttt{server\_context.metrics} dictionary under the corresponding client identifier and round number (e.g., \texttt{server\_context.metrics[client\_id][round]}). 
This allows the server to maintain a structured record of client performance over rounds while keeping client data isolated. The aggregated metrics can then be used for centralized logging or to enable adaptive aggregation strategies, such as weighting client updates based on their reported performance. This design preserves the simplicity of the hook-based system while improving transparency and extensibility in experiment tracking.

Furthermore, this hook system can enable more advanced algorithmic development for FL systems. For example, users can easily implement the FedCostAware 
\cite{sinha2025fedcostaware} scheduling strategy. 
Using FedCostAware the server can suggest a certain client to turn off their computing resources (i.e., the process running the client agent) if the server estimates that the client will be idle for a long time. To achieve so, a user could implement a strategy where the server, using the \texttt{before\_client\_selection} hook, calculates the estimated finish time for the slowest client in the round and sends this estimate in \texttt{server\_context} as metadata. Then, on the client side, a function registered to the \texttt{after\_local\_train} hook can make an intelligent decision: Upon completing its local training, it can compare its finish time to the server's estimate for the end of the round. If the remaining idle time is substantial enough to outweigh the cost of instance termination and subsequent spin-up, the client can autonomously decide to shut down its compute instance, thereby minimizing operational costs. This advanced use case, combining both server-side and client-side hooks, demonstrates how our architecture can be extended to support complex algorithm implementations, directly supporting our goal of balancing ease of use with research flexibility.

\begin{lstlisting}[
    style=PythonStyle,
    label={lst:example1},
    caption={An example for client evaluation hook with metric tracking.},
    float,
    floatplacement=t
]
@on_event('after_local_train')
def eval_local(server_context, client_context):
    # Evaluate model on client's test data.
    model = client_context.model
    data  = client_context.data.test_loader
    loss, acc = evaluate(model, data)
    cid, rnd = client_context.id, server_context.round

    # Store metrics in server context
    server_context.metrics[cid][rnd] = {'test_loss': loss, 'test_acc': acc}
\end{lstlisting}

\begin{lstlisting}[
    style=PythonStyle,
    label={lst:example2},
    caption={An example for implementing server–client coordination for cost-aware and intelligent client machine shutdown.},
    float,
    floatplacement=t
]
@on_event('before_client_selection')
def set_round_eta(server_context):
    # Server predicts round finish time and shares ETA.
    eta = max(c.expected_finish for c in server_context.clients)
    server_context.set_metadata('round_eta', eta)

@on_event('after_local_train')
def check_idletime_and_shutdown(server_context, client_context):
    # Client decides on shutdown based on ETA and cost model.
    eta = server_context.get_metadata('round_eta')
    idle = max(0, eta - time.time() - client_context.spin_up_time)
    if idle > client_context.shutdown_threshold:
        client_context.terminate_self()
\end{lstlisting}

\subsection{Providing FL as a Service}
\begin{figure}
    \centering
    \includegraphics[width=1\linewidth]{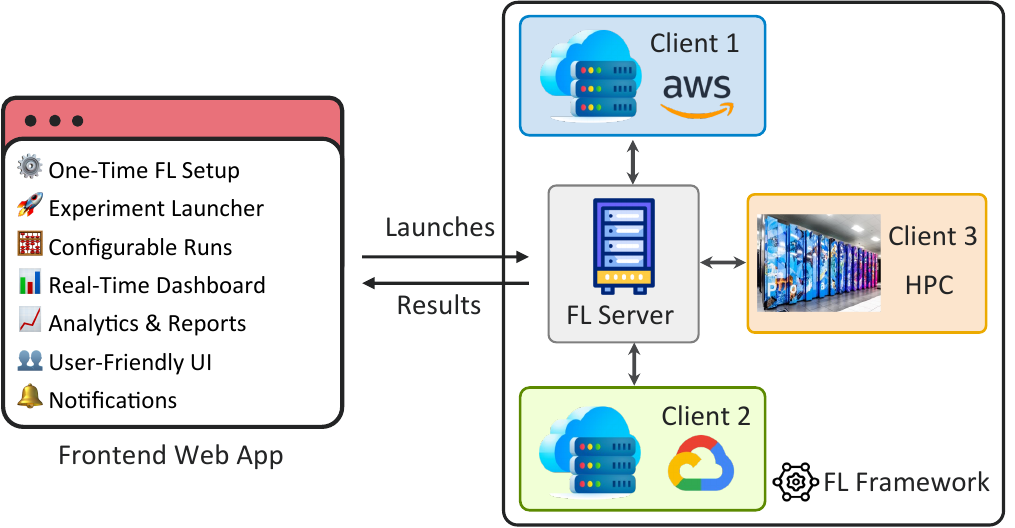}
    \caption{\textbf{Federated Learning as a Service (FLaaS).} A unified web platform enabling one-time client setup, automated experiment management, real-time monitoring, and analytics across heterogeneous clients.}
    \label{fig:flass}
\end{figure}

To further democratize access to cutting-edge FL capabilities and lower the overall technical barriers to its adoption, a good FL framework implementation should be extended to offer FL as a Service (FLaaS), as illustrated in Figure~\ref{fig:flass}. Such a service-oriented paradigm would further make FL accessible to a broader community of researchers, enterprises, and scientific institutions that may lack the engineering capacity to manage multiple complex distributed training runs. With the service, clients would only need to perform a one-time setup to register and configure their local computing environments. Once integrated, the service would handle the secure federation setup, orchestration, and management of FL experiments, automating many of the tedious and error-prone steps typically involved in distributed training. By abstracting away infrastructure-level complexities, the system enables practitioners to focus on model development and experimentation rather than on deployment logistics.

\begin{table*}[htbp]
\centering
\caption{Comparison of popular open-source federated learning frameworks across key enterprise-level capabilities.\newline  \cmark: Supported, \pmark: Partially Supported, \xmark: Not Supported}
\renewcommand{\arraystretch}{1.1}
\setlength{\tabcolsep}{4pt}
\begin{tabular}{lcccccccc}
\hline
\textbf{Capability} & \textbf{FederatedScope} & \textbf{NVFlare} & \textbf{OpenFL} & \textbf{FedScale} & \textbf{FedLab} & \textbf{Flower} & \textbf{FedML} & \textbf{APPFL} \\
\hline
Scalable Local Simulation & \cmark & \cmark & \pmark & \cmark & \cmark & \cmark & \cmark & \cmark \\
Seamless Simulation/Deployment Transition & \pmark & \cmark & \cmark & \pmark & \xmark & \cmark & \cmark & \pmark \\
Heterogeneous Deployment & \pmark & \pmark & \cmark & \cmark & \pmark & \cmark & \cmark & \cmark \\
Hierarchical Abstractions & \cmark & \cmark & \pmark & \cmark & \cmark & \cmark & \cmark & \cmark \\
Privacy \& Security Integration & \cmark & \cmark & \cmark & \xmark & \xmark & \pmark & \pmark & \cmark \\
\hline
\end{tabular}
\label{tab:framework_capabilities}
\end{table*}
A well-designed service platform could offer fire-and-forget experiment management, allowing practitioners to launch, monitor, and reproduce FL experiments across heterogeneous clients through a unified interface. For example, practitioners could easily configure and execute multiple experiment runs with varying hyperparameters, resource allocation settings, or client selections via a web-based frontend, without needing to manually modify code or deployment scripts. Additionally, the service could provide built-in visualization and analytics tools for real-time monitoring and post-experiment analysis. Interactive dashboards could display the overall experiment settings, model convergence trends, client participation statistics, communication overhead, and resource utilization. Automated reporting features could further summarize experiment outcomes, enabling reproducible benchmarking and performance comparison across different FL algorithms or configurations.

\section{Review of Existing FL Frameworks}

Numerous open-source FL frameworks have emerged to enable distributed and privacy-preserving model training. These systems advance FL research by supporting scalable simulation, modular APIs, and deployment in real-world settings. However, no solution has yet been offered that covers all required capabilities proposed in this paper for an ideal enterprise-level FL framework to seamlessly integrate local prototyping, large-scale experimentation, and heterogeneous deployment. As summarized in Table~\ref{tab:framework_capabilities}, most frameworks provide only partial coverage, lacking full integration of scalability, privacy, and smooth transition from simulation to production.

FederatedScope \cite{xie2022federatedscope} adopts a flexible, event-driven architecture that supports heterogeneous and asynchronous simulations, making it well-suited for research prototyping. However, its deployment remains largely confined to experimental settings. NVFlare \cite{roth2022nvidia} offers one of the most complete ecosystems, combining scalable simulation, seamless research-to-production transition, and robust privacy mechanisms, though its focus remains mainly cloud-centric with limited HPC adaptability. OpenFL \cite{foley2022openfl} targets production-grade cross-silo deployments with strong privacy guarantees, but provides limited scalability for simulation and flexibility for algorithmic exploration. FedScale \cite{lai2022fedscale} delivers a powerful benchmarking suite for large-scale heterogeneous FL experiments yet lacks orchestration tools for deployment and security protection techniques. FedLab \cite{zeng2023fedlab} is a lightweight simulator optimized for communication-efficient testing but offers no built-in support for distributed or secure deployment and safety guarantees. Flower \cite{beutel2020flower} and FedML \cite{he2020fedml} emphasize scalability and flexibility across cloud, edge, and mobile platforms, supporting smooth migration from simulation to deployment. Nonetheless, their security mechanisms are optional or externally integrated. APPFL \cite{li2025advances} distinguishes itself through modular extensibility and integrated privacy via differential privacy, secure aggregation, and Globus authentication, though it lacks native support for enterprise orchestration platforms such as Kubernetes.

\section{Conclusion}
In this paper, we present our vision for an enterprise-level, privacy-preserving federated learning (FL) framework for bridging the gap between rapid research prototype to secure, large-scale experiment deployment. By analyzing the typical end-to-end workflow of FL experiments, we identify several key capabilities required for such a framework, including scalable local simulation, seamless simulation and deployment transition, adaptability across heterogeneous computing infrastructures, hierarchical abstractions for usability and extensibility, and comprehensive privacy guarantees. We also propose a framework architecture design that decouples the FL-related core logic from the complex communication and resource orchestration, with event-driven hooks to enhance extensibility. We further emphasize the potential of extending the FL framework into a hosted service (FL as a Service) to lower the FL adoption barriers and promote broader accessibility. Together, these design principles outline a path towards a scalable, secure, and user-friendly FL system that can power the next generation of collaborative and privacy-preserving AI for scientific discovery. 

\section*{Acknowledgments}

This work was supported by the U.S. Department of Energy, Office of Science, Advanced Scientific Computing Research, under Contract DE-AC02-06CH11357.

\clearpage
\bibliographystyle{IEEEtran}
\bibliography{PPFL}

\end{document}